\documentstyle[epsfig]{mn}

\def\Msun{\ifmmode{~{\rm M}_\odot}\else${\rm M}_\odot$~\fi}
\def\kms{\ifmmode{$~km\thinspace s$^{-1}}\else km\thinspace s$^{-1}$\fi}

\def\ee{\end{equation}}
\def\be{\begin{equation}}

\title{White dwarfs and Galactic dark matter}

\author[Chris Flynn, Janne Holopainen and Johan Holmberg] 
{Chris Flynn$^{1,2}$, Janne Holopainen$^1$ and Johan Holmberg$^{1,3}$\\
$^1$Tuorla Observatory, Piikki\"o, FIN-21500, Finland\\ 
$^2$Centre for Astrophysics and Supercomputing, Swinburne University of 
Technology, Hawthorn, Australia\\
$^3$Lund Observatory, Box 43, SE-22100, Lund, Sweden}

\date{}

\begin{document}

\maketitle

\voffset=-1.0cm     

\begin{abstract} 
We discuss the recent discovery by Oppenheimer et al (2001) of old,
cool white dwarf stars, which may be the first direct detection of
Galactic halo dark matter.  We argue here that the contribution of
more mundane white dwarfs of the stellar halo and thick disk would
contribute sufficiently to explain the new high velocity white dwarfs
without invoking putative white dwarfs of the dark halo. This by no
means rules out that the dark matter has been found, but it does
constrain the overall contribution by white dwarfs brighter than $M_V
\approx 16$ to significantly less than 1\% of the Galactic dark
matter. This work confirms a similar study by Reyl\'e et al (2001).
\end{abstract}


\section{Galactic Dark Matter as Faint Stars?}

The MACHO and EROS microlensing projects have detected a microlensing
signal from dark objects in the Galactic halo (Alcock et al. (2000)
(The MACHO collaboration), and Lassarre et al (2000) (The EROS
Collaboration). Taken together, their results show the microlensing
can be explained by a Galactic dark halo 20\% of which is in the form
of approximately 0.5 M$_\odot$ objects. This suggests that the
responsible objects could be low mass main sequence red dwarf stars,
or white dwarf stars, but in order to have escaped detection to date,
they must be very faint.

Red dwarfs are too luminous, or they would have been detected directly
in the Hubble Deep Field (HDF) (Flynn et al 1996, Elson et al 1996 and
Mendez et al 1996).  White dwarfs were considered more difficult to
rule out directly, until the surprising discovery was made that white
dwarfs which have had long enough to cool, cease becoming fainter and
redder, but remain at approximately constant luminosity while
becomeing {\it bluer}, due to the development of H$_2$ in their
atmospheres which induce very non-blackbody spectra (Hansen, 1999a,
1999b).

\begin{figure}
\begin{center}
\epsfig{file=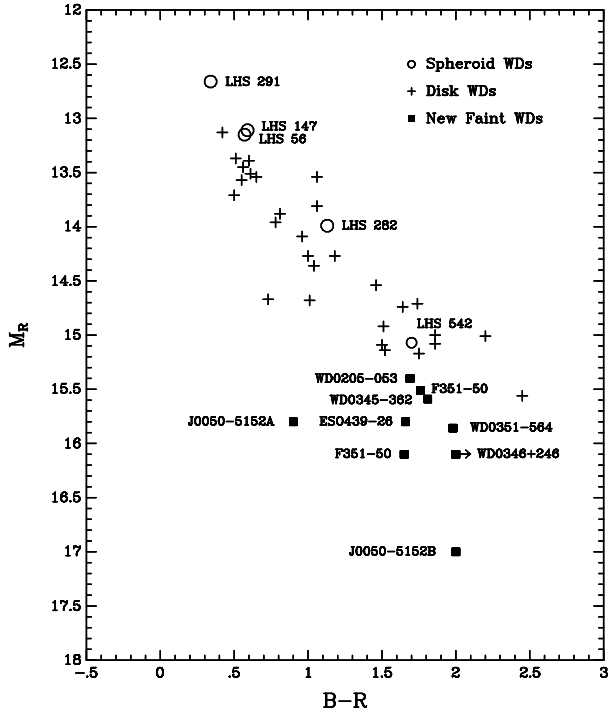,width=80mm}
\caption[]{Colour ($B-R$) versus absolute magnitude (in the $R$-band) diagram
of nearby white dwarfs. Disk white dwarfs are shown by crosses, and spheroid
white dwarfs as circles. Until recently, the faintest white dwarfs known were
at $M_R \approx 16$, but recently four white dwarfs have turned up in new
surveys designed to probe for them (squares). These objects are shown as
triangles. One object, J0050-5152, is a binary, the secondary being much
fainter and its status as a white dwarf is still to be confirmed. All of these
new white dwarfs are still relatively bright, and probably not faint enough to
be good dark matter candidates.}
\label{cmd}
\end{center}
\end{figure}

The possibility that very faint, blue objects had been found arose
when the HDF was imaged at a second epoch and analysed by Ibata et al
(2000). The number of moving objects, their colours, magnitudes and
proper motions were all consistent with the detection of a significant
fraction of the dark halo matter in the form of old, cool white
dwarfs.  However, a third epoch observation of the HDF did not confirm
the proper motions of the objects (Richer, 2001), but the idea that
the dark matter had been detected had already spurred many groups to
search for local counterparts --- with some degree of success.  As a
result, a number of very low luminosity white dwarfs have turned up in
new proper motion studies by Ibata et al (2000), Hodgkin et al (2000),
Scholz et al (2000), Goldman (2000), de Jong et al (2000) and
Oppenheimer et al (2001), along with a very low luminosity white dwarf
identified by Ruiz et al (1995), now viewed with new significance.  A
colour magnitude diagram for these objects in shown in Figure
\ref{cmd}.  The recently discovered white dwarfs are shown as
triangles --- they are all fainter than the end of the white dwarf
cooling sequence at $M_R \approx 15.5$, and have velocities typical of
the spheroid white dwarfs.

Oppenheimer et al (2001 --- hereafter OHDHS) have conducted the largest of
these recent surveys. They discovered 38 high velocity white dwarfs and derived
a space density for their objects which corresponds to approximately 2\% of the
Galactic dark halo density at the Sun (aproximately 0.01 \Msun pc$^{-3}$). This
is a small but significant fraction of the dark halo density, not high enough
to explain the microlensing events (which require that about 20\% of the dark
halo be in the form of $\approx 0.5$ \Msun objects, but still significantly
higher than the expected contribution of white dwarfs from all the well
understood Galactic stellar populations --- disk, thick disk, and stellar halo
(or spheroid).

A very similar survey to OHDHS, in terms of the local volume of space surveyed
for high proper motion stars, is the Luyten Half Second catalog, Luyten
(1979). Flynn et al (2001) have searched the LHS and two other older proper
motion surveys for nearby dark halo white dwarf candidates.  The LHS covers
more than half the sky, has a limiting magnitude of $V=18.4$, with proper
motions in the range 0.2 to 0.5 arcsec/year.  A recent independent analysis
(Monet et al 2000) shows that the LHS is substantially (90\%) complete within
these limits, based on a new, deeper survey over a small area within the LHS.

Gibson and Flynn (2001) have searched the LHS for objects of the type detected
by OHDHS, but found at most a few, even though the survey covers a similar
volume. If 2\% of the local dark matter is composed of white dwarfs, then a few
tens of objects had been expected in the LHS, whereas Flynn et al (2001) had
earlier analysed the LHS and two other proper motion surveys in detail, and had
found no convincing evidence that any of the high proper motion objects could
be associated with ``dark halo white dwarfs''. All the objects were broadly
consistent with coming from the visible Galactic populations.

An alternative explanation for the OHDHS white dwarfs is that they are from
existing Galactic populations, and do not represent a new population from the
dark halo. That they are members of the thick disk has been argued by Reid et
al (2001) and Hansen (2001). We also argue for this view in this paper by
modelling the expected numbers of white dwarfs which would be found in proper
motion surveys from the known Galactic populations. Our work supports a
similar, independent study by Reyl\'e et al (2001).

\section{Model of low luminosity stars in the Solar neighbourhood}

We have built a model of the low mass stellar content of the Solar
neighbourhood, including the luminosity function, density and kinematics, in
order to simulate actual proper motion surveys via a Monte-Carlo technique. The
model allows us to predict the expected number of low mass stars which would be
recovered in a proper motion survey directed toward any point on the sky,
covering a given area, with a given apparent magnitude limit and a detection
window of proper motions.  For example, the LHS catalog covers a little more
than half the sky centered on the Northern hemisphere, has an apparent
magnitude limit of $R = 18.6$, and recovered proper motions $\mu$ in the window
$0.5 < \mu < 2.5$ arcseconds per year.

We model a sphere centered on the Sun with a radius of up to 288 pc, which is
sufficiently distant from the Sun for all the surveys we consider. Only low
luminosity ($M_V > 12.5)$ stars are included in the model.

\subsection{Populations in the Model}

The local Galactic components represented in the model are the disk, the thick
disk, the stellar halo and the dark halo.

\begin{itemize}
\item[$\bullet$] {\it The disk} component consists of M dwarfs and white
dwarfs. The M dwarfs are drawn from a luminosity function shown in Table
\ref{MLF}, which has been measured from faint star counts with HST (Zheng et al
2001). The white dwarfs are drawn from the luminosity function shown in Table
\ref{WDLF}, which comes from Liebert et al (1988).

\begin{table}
\begin{center}
\caption{Adopted Luminosity function for disk M dwarfs}
\begin{tabular}{cccc}
\hline
$M_{V}$ & log $\Phi{}(M_{V})$  & $M_{V}$ & log $\Phi{}(M_{V})$  \\
        & stars pc$^{-3} M_{V}^{-1}$    &&  stars pc$^{-3} M_{V}^{-1}$ \\
\hline
10 & $-$2.12 & 14   & $-$2.30 \\
11 & $-$1.92 & 15.5 & $-$2.67 \\
12 & $-$1.89 & 17.5 & $-$2.61 \\
13 & $-$2.19 &      &         \\
\hline
\end{tabular}
\label{MLF}
\end{center}
\end{table}

\begin{table}
\caption{Adopted Luminosity function for disk white dwarfs}
\begin{center}
\begin{tabular}{cccc}
\hline
$M_{V}$ & log $\Phi{}(M_{V})$ & $M_{V}$ & log $\Phi{}(M_{V})$ \\
        & stars pc$^{-3} M_{V}^{-1}$ &  & stars pc$^{-3} M_{V}^{-1}$ \\
\hline
11.0 & $-$4.02 & 14.0 & $-$2.93 \\
11.5 & $-$3.92 & 14.5 & $-$3.03 \\
12.0 & $-$3.82 & 15.0 & $-$2.98 \\
12.5 & $-$3.54 & 15.5 & $-$3.09 \\
13.0 & $-$3.22 & 16.0 & $-$4.14 \\
13.5 & $-$3.06 & 16.5 & $-$4.50 \\
\hline
\end{tabular}
\label{WDLF}
\end{center}
\end{table}

The disk white dwarfs have velocity dispersion components $\sigma = (\sigma_U,
\sigma_V, \sigma_W)$ (where $U, V, W$ are the usual space velocities in the
directions of the Galactic center, Galactic rotation and perpendicular to the
Galactic plane) of $\sigma = (40, 30, 20)$ \kms, and a mean motion (asymetric
drift) relative to the Sun of $-20$ \kms. For the disk M dwarfs, we adopt
velocity dispersion components of $\sigma = (35, 25, 15)$ \kms, and a
asymetric drift, $V_{\mathrm ass} = 15$ \kms. These values are slightly lower
than for the white dwarfs, because the mean age of the M dwarfs is certainly
lower than the white dwarfs. The young M dwarf component is measured to have
$\sigma = (30, 17, 12)$ \kms, and the old M dwarf component $\sigma = (56, 34,
31)$ \kms, in an analysis of the Hipparcos results by Upgren et al (1997).  In
both the surveys analysed in this paper, the number of M dwarfs recovered from
the proper motion windows is quite sensitive to the kinematics adopted; our
values have been choosen to produce about the right number of M dwarfs. Note
though thata our main interest is to predict the number of white dwarfs in the
proper motion surveys, while the M dwarfs are only included as a consistency
check on the modelling. The number of white dwarfs predicted in the surveys is
less sensitive to the adopted kinematics, because the white dwarfs are
generally closer than the M dwarfs.

\item[$\bullet$]{\it The thick disk} component of the model consists of white
dwarfs selected from the same luminosity function as the disk white dwarfs, but
with space density of 5\% of the disk (Reyl\'e and Robin, 2001). This is
consistent with recent estimates of the thick disk density as between 2\% and
10\% of the disk density (Kerber et al 2001). We adopt an uncertainty in the
thick disk local normalisation of about a factor of two.  The thick disk stars
are given velocity dispersion components $\sigma = (80, 60, 40)$ \kms, and an
asymetric drift $V_{\mathrm ass} = 40$ \kms (see e.g. Morrison et al 1990).
We also include thick disk M dwarfs, selected from the same luminosity function
as the disk M dwarfs, with a normalisation of 5\% of the disk, and with the
same kinematic parameters as the thick disk white dwarfs.

The significantly larger space motions of the thick disk stars means that they
are several times more likely to be found in a typical proper motion survey
than disk stars. Reyl\'e et al (2001) model the thick disk in their study of
white dwarf proper motions as $\sigma = (67, 51, 42)$ \kms and $V_{\mathrm ass}
= 53$ \kms. We have found that adopting either ours or Reyl\'e et al's
kinematics for the thick disk ends up giving quite similar results, i.e. the
model predictions are not very sensitive to the adopted thick disk kinematics.

\item[$\bullet$]{\it The stellar halo}, (or spheroid) part of the
model consists of white dwarfs drawn from a luminosity function due to
Liebert (2001, private communication). The luminosity function is
based on 7 white dwarfs, identified as members of the ``halo'' on the
basis of a high tangential velocity, $V_{\mathrm tan} > 160$ \kms
obtained as part of an ongoing analysis of all the faint stars in the
LHS. The luminosity function is shown in Table \ref{shalowd}. Note
that we have converted bolometric magnitudes to V-band magnitudes
using Eqn 1 of Liebert et al (1988). The total number density in these
objects is $3.2 \times 10^{-5}$ stars pc$^{-3}$ which corresponds to a
mass density of $1.9 \times 10^{-5}$ \Msun pc$^{-3}$ for a white dwarf
mass of 0.6 \Msun. This is approximately 15\% of the local mass
density of the stellar halo (as seen in subdwarfs) of $1.5 \times
10^{-4}$ \Msun pc$^{-3}$ (Fuchs and Jahrei\ss, 1997).

\begin{table}
\begin{center}
\caption{Adopted Luminosity function for stellar halo white dwarfs}
\begin{tabular}{ccc}
\hline
$M_{\mathrm bol}$ & $M_{V}$ & log $\Phi{}(M_{V})$  \\
                  &         &  stars pc$^{-3} M_{V}^{-1}$ \\
\hline
11.75 & 12.10 & $-$6.15 \\
14.85 & 15.25 & $-$4.50  \\
\hline
\end{tabular}
\label{shalowd}
\end{center}
\end{table}

For the stellar halo we adopt velocity dispersion components of $\sigma = (141,
106, 94)$ \kms (Chiba and Beers, 2001) and an asymetric drift of $180$ \kms
(i.e. there is a small net rotation in the Galactocentric coordinate frame).
The adopted values have very little impact on the conclusions of the paper,
because stars with such high velocities populate the proper motion selection
window quite well. For example, changing these values to $\sigma = (131, 106,
85)$ \kms and an asymetric drift of $229$ \kms (as used by Reyl\'e et al 2001)
only changes the predicted number of white dwarfs by a few percent.

We further include stellar halo M dwarfs, drawn from the disk M dwarf
luminosity function, but with a local density reduced by a factor of 500 (see
e.g. Morrison 1993). They are assigned the same kinematics as the white dwarf
stellar halo.

\item[$\bullet$]{\it The dark halo} part of the model consists of white dwarfs
only. We adopt velocity dispersion components of $\sigma = (156, 156, 156)$
\kms{} and an asymetric drift of $V_{\mathrm as} = 220$ \kms, which simulates
an isothermal population with a density falloff with Galactic radius of $\rho
\propto R^{-2}$, i.e. a sufficient but not necessary condition to explain the
flat Galactic rotation curve. Our starting point for the mass density of these
stars is 2\% of the local dark halo density of 0.008 \Msun pc$^{-3}$ (Gates et
al 1998) with an average WD mass of 0.6 \Msun, as adopted in the OHDHS survey.

We have adopted a very simple luminosity functions for the dark halo white
dwarfs.  We give all the dark halo white dwarfs the same absolute magnitude,
$M_V = 15.9$, which is the faintest absolute magnitude of any of the putative
dark halo white dwarfs found in the OHDHS survey. We show that this choice,
combined with a 2\% dark halo, leads to significantly larger numbers of dark
halo white dwarfs in both surveys than actually observed.  Any other choice of
luminosity function which is still consistent with the WDs in the OHDHS sample
(i.e. all stars are brighter than $M_V = 15.9$) would produce an even greater
disagreement between the simulations and the LHS and the number of recovered
WDs in the OHDHS sample.

\end{itemize}

\begin{table}
\begin{center}
\caption{Kinematic parameters of populations in the model}
\begin{tabular}{lcc}
\hline
Population & ($\sigma_U, \sigma_V, \sigma_W)$  & $V_{\mathrm as}$ \\
           & \kms                              & \kms \\
\hline
Disk M dwarfs      &$  (35,25,15)  $ & $15$ \\ 
Disk white dwarfs  &$  (40,30,20)  $ & $20$ \\ 
Thick disk         &$  (80,60,40)  $ & $40$ \\ 
Stellar halo       &$ (141,106,94)$  & $180$ \\ 
Dark halo          &$ (156,156,156)$ & $220$ \\ 
\hline
\end{tabular}
\label{pops}
\end{center}
\end{table}

The {\it dark halo white dwarfs} share similar kinematic properties with the
{\it stellar halo white dwarfs} (i.e. high random velocities), but differ from
them in a key respect worth pointing out (although it has no effect on the
modelling or conclusion in this paper).  The density distribution of the {\it
stellar halo} is well determined by luminous stars, and follows a power law
outward from the Galactic center, $\rho(R) \propto R^{-3.5}$, where $R$ is the
Galactocentric radius (see e.g. Wetterer and McGraw 1996). If the dark halo
were made entirely of dim white dwarfs, they would require a density
distribution which follows $\rho(R) \propto R^{-2}$, in order to generate a
gravitational field which would account for the observed flat rotation curves
of disk galaxies. However, we know from the microlensing results that they do
not dominate the total mass of the halo.  In that case, they could still have
an $R^{-3.5}$ distrubution (as discussed by Gates and Gyuk 2001).  In their
model, ``normal'' Cold Dark Matter makes up the rest of the mass distribution
and does follow an $R^{-2}$ distribution.  For the purposes of the modelling
here, the local volume surveyed is so small that the density of the white
dwarfs over the volume is essentially constant. 

The kinematic properties of all the model populations are shown in Table
\ref{pops}.

\section{Modelling the LHS and OHDHS surveys}

The model allows us to run Monte-Carlo simulations of a proper motion
survey. Stars are generated within a small sphere around the Sun, with a
uniform density distribution in the case of the halo and thick disk
populations, and with a density distribution which is falling off in $z$ as
sech$^2(z/z_h)$ in the case of the disk (where we adopt $z_h = 125$ pc, which
is equivalent to an exponentially falling disk scale height of 250 pc far from
the plane).

For each star a $V$-band absolute magnitude is selected from the appropriate
luminosity function, and its apparent magnitude computed. A $V-I$ colour is
assigned to the stars as follows:

For the disk M dwarfs we use the empirically calibrated relation due to Reid
(1991)

\begin{displaymath}
(V - I)_{MD} = 0.297 \times M_V - 0.858
\end{displaymath}

For the white dwarfs, $V - I$ colors are calculated similarly by a relation:

\begin{displaymath}
(V - I)_{WD} = 0.385 \times M_V - 4.85
\end{displaymath}

This relation comes from fitting white dwarfs in the sample of Bergeron et al
(2001). We have also derived transformation equations for white dwarfs from
$V-I$ to other colours from the same sample:

\begin{displaymath}
B - V = 1.05 \times (V-I) - 0.13
\end{displaymath}

\begin{displaymath}
R - I = 0.473 \times (V-I) + 0.01
\end{displaymath}

Note that a small Gaussian random number with standard deviation $\sigma =
0.05$ is added to the $V-I$ colour to avoid clutter in the figures.

\begin{figure}
\begin{center}
\epsfig{file=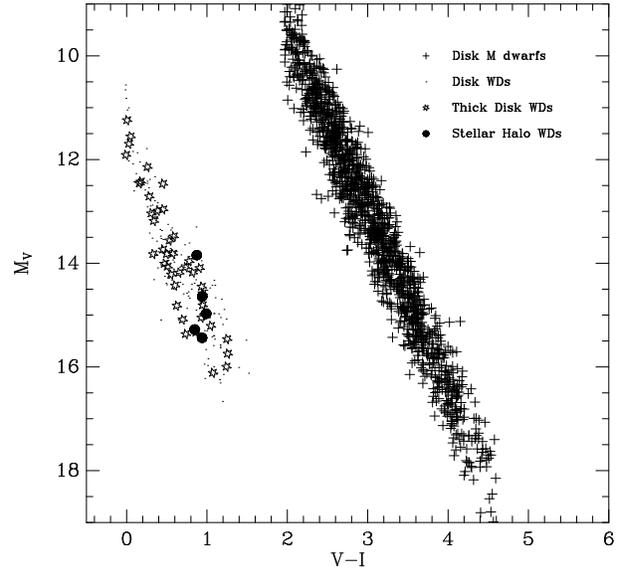,width=80mm}
\caption[]{The main components of the model in the $M_V$ versus $V-I$ colour
magnitude diagram. A small random number is added after assigning the color to
avoid stars settling strictly on a line. The main sequence stars and white
dwarfs are clearly separated in this plane.}
\label{mvvi}
\end{center}
\end{figure}

An absolute magnitude $M_V$ versus colour $V-I$ diagram from a typical
simulation of the LHS catalog is shown in Figure \ref{mvvi}. Main sequence disk
stars dominate the sequence on the right side. The left sequence consists of
white dwarfs of the disk, the thick disk and the stellar halo.

\subsection{Simulation of the OHDHS survey}

The OHDHS survey covers 4165 square degrees, centered on the South Galactic
Pole, to a limiting magnitude of $R_{59F} = 19.7$ in a proper motion $\mu$
window of $0.33 < \mu < 3.0$ arcsecs/year. In order to simulate the OHDHS
sample, we generate colour magnitude diagrams of nearby stars using the model
described in the previous section. We then transform the $M_V$ and $V-I$ values
to the bands used in the OHDHS survey.  For {\it white dwarfs}, we use the
following transformations to the (photographic) $R_{59F}$ and the
(photographic) $B_J$ bands (i.e. for the III-aJ emulsion) used in the OHDHS
survey:

\begin{eqnarray*}
R - R_{59F} = & 0.006 + 0.059 \times (R-I) - 0.112 \times (R-I)^2 \\ & - 0.0238
              \times (R-I)^3
\end{eqnarray*}

\begin{displaymath}
B - B_J = 0.28 \times (B-V) ~~~~~~~~{\rm for}~ -0.1 \le B-V \le 1.6
\end{displaymath}

\noindent which come from Bessell (1985) and Blair and Gilmore (1982).

For the {\it M dwarfs} we have derived a transformation from $R-I$ to
$B_J-R_{59F}$ based on Figure 10 of Hambly et al (2001), who describe in detail
the Super COSMOS sky survey, upon which the OHDHS survey is based.

\begin{figure}
\begin{center}
\epsfig{file=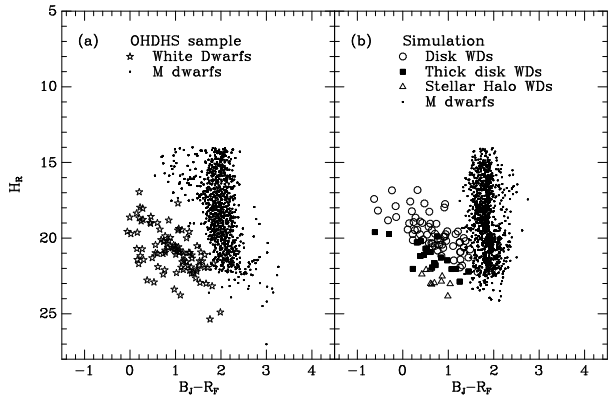,width=80mm}
\caption[]{Reduced proper motion versus colour plots for the OHDHS and
simulated samples. Panel (a) shows the data from OHDHS, with white dwarfs
marked by stars and M dwarfs by dots. Panel (b) shows a typical simulation of
the OHDHS survey using our model. The main features of the data are well
reproduced by the simulations: most of the white dwarfs detected are plausibly
from the disk, thick disk and stellar halo. Two of the OHDHS white dwarfs (at
high reduced proper motion) are not easily accounted for by the model, and are
very interesting as dark matter candidates.}
\label{oppcomp}
\end{center}
\end{figure}

Figure \ref{oppcomp} shows the results of a simulation in the reduced proper
motion $H_R$ versus colour $B_J-R$ plane (hereafter, $H_R$ and the $R$-band
refers to the $R_{59F}$ band, used by OHDHS). Panel (a) shows the simulation,
and panel (b) shows the actual OHDHS data.  Circles mark disk stars, squares
mark thick disk stars and triangles stellar halo stars. There is good agreement
between the simulated sample and the actual data. Firstly, the disk white
dwarfs form a wide sequence from $(B_J-R,H_R) = (0.0,17.0)$ to $(B_J-R,H_R) =
(1.5,22.0)$.  Most white dwarfs appear to be members of the disk. Below this
sequence, most of the thick disk and stellar halo white dwarfs appear, because
they have generally greater space velocities than the disk stars, and thus
higher reduced proper motions.

The M dwarfs lie in an almost vertical line at $B_J-R \approx 2.0$ in both
panels.  The total number of M dwarfs in the simulation is similar to the
observations, as a result of adjusting the kinematic parameters of the disk to
acheive this consistency. The number of disk M dwarfs turns out to be quite
sensitive to the disk kinematics, and we acheived this good fit by using the
kinematic parameters for the disk shown in Table \ref{pops}. The kinematic
parameters adopted are however quite consistent with the observed kinematics of
M dwarfs (Upgren et al (1997), and also section 2.1).

To a limiting magnitude of $R = 19.7$, OHDHS identify a total of 97 white
dwarfs, of which 82 have direct spectroscopic confirmation, and 15 are assumed
with reasonable confidence to be white dwarfs based on their position in the
reduced proper motion versus colour diagram.

We have computed the expected numbers of various white dwarf types using the
model. The expected numbers are: for the disk $60\pm8$ WDs, for the thick disk
$21\pm5$ Wds and for the stellar halo, $10\pm3$ WDs, for a total of 91 WDs.

The adopted kinematical parameters have a small effect on the predicted numbers
of white dwarfs. For example, adopting the disk, thick disk and halo kinematics
used by Reyl\'e et al (2001), (i.e. for the disk: $\sigma = (42.1, 27.2, 17.2)$
\kms, and $V_{\mathrm as} = 16.6$ \kms, for the thick disk $\sigma = (67, 51,
42)$ \kms, and $V_{\mathrm as} = 53$ \kms, and for the halo $\sigma = (131,
106, 85)$ \kms, and $V_{\mathrm as} = 229$ \kms, compare with Table
\ref{pops}), we obtain 54 disk WDs, 18 thick disk WDs and 11 halo WDs, that is
the numbers are very similar.

The total number of white dwarfs in the model and the OHDHS sample are in good
agreement. We now check the relative numbers of white dwarfs of each population
type. The simulations indicate that a neat dividing line between disk and other
types of white dwarfs can be drawn from $(B_J-R,H_R) = (-0.5,17.0)$ to
$(B_J-R,H_R) = (2.0,26.0)$. Counting white dwarfs below this line in the
simulations yields $27 \pm 5$ stars, compared to 21 white dwarfs in the OHDHS
sample.

Oppenheimer et al (2001) argue that many or all of these higher reduced proper
motion white dwarfs are part of a new dark halo population, comprising some 2\%
of the local dark matter halo density. We argue instead that these white dwarfs
can just as well be interpreted as coming from a mixture of the thick disk and
stellar halo. Further evidence for this assertion comes from the kinematics of
the white dwarfs.

\begin{figure}
\begin{center}
\epsfig{file=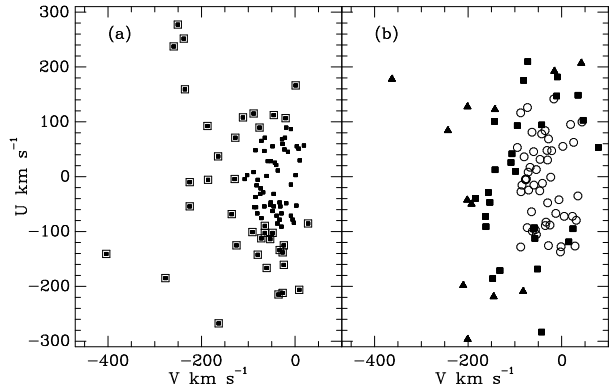,width=80mm}
\caption[]{Kinematics of the white dwarfs. Panel (a) shows the $V$
versus the $U$ velocities for the white dwarfs in the Oppenheimer et
al sample. Panel (b) shows results from a typical simulation using
disk, thick disk and stellar halo white dwarfs, but no dark halo white
dwarfs. (The same symbols are used as in Figure \ref{oppcomp}(b)). The
main features of the observations appear to be well reproduced by the
classical stellar populations alone.}
\label{Oppkin}
\end{center}
\end{figure}

We show in Figure \ref{Oppkin} the space velocities of the white
dwarfs in the $V$ versus $U$ plane, i.e. projected onto the Galactic
plane. A typical simulation is shown in panel (a) and the OHDHS sample
stars in panel (b). In the simulation, disk stars are marked by
diamonds thick disk stars by crosses and halo stars by circles.  The
similarity between the diagrams is striking. In particular, the thick
disk and stellar halo white dwarfs in the simulation lie mostly
outside the 2-$\sigma$ circles used by Oppenheimer et al (2001) to
isolate the dark halo stars (squares in squares). (The 2-$\sigma$
circle is where stars with velocities twice that of the disk velocity
dispersion lie, relative to the mean motion of the old disk at $(V,U)
= (-35, 0)$ \kms). In a typical simulation we count $41 \pm 6$ white
dwarfs outside the 2-$\sigma$ circle, while OHDHS find 37 such
stars. The distribution of these simulated stars in the $(V,U)$ plane
is also found to be very similar to the observations.

\begin{figure}
\begin{center}
\epsfig{file=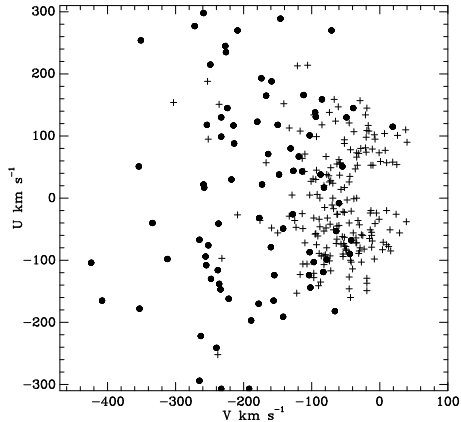,width=60mm}
\caption[]{Kinematics of a sample of nearby dwarfs (Fuchs and
Jahrei\ss, private communication 2002). Crosses indicate stars in the
range $-1.6 <$ [Fe/H] $< -0.5$ (broadly, ``thick disk'' stars) while
circles are stars with [Fe/H] $<-1.6$ (broadly, ``halo'' stars). The
distribution of velocities for the two metallicity types is very
similar to that of the white dwarfs of the OHDHS sample (Figure
\ref{Oppkin}(a).}
\label{subdw}
\end{center}
\end{figure}

We show in Figure \ref{subdw} the $V$ versus $U$ velocities for a
sample of local dwarf stars for which [Fe/H] is available (Fuchs and
Jahrei\ss, private communication 2001). The availability of abundances
means that the stars can be classified by their population type
directly, rather than compared statistically to the models (as is the
case for the white dwarfs). There are 282 stars in total, of which 245
are in the metallicity range $-1.6 <$ [Fe/H] $< -0.5$, and are here
termed ``thick disk'' stars, while 37 have [Fe/H] $< -1.6$ and are
here termed ``halo'' stars. What is striking about this figure is how
similarly the $(V,U)$ distribution of the stars is to the OHDHS samlpe
(Figure \ref{Oppkin}(a)). Thick disk stars dominate the region around
$(V,U) = (-35, 0)$ \kms), while the halo stars fill a much broader
region centered roughly at $(V,U) = (-220, 0)$ \kms). 

\subsection{Simulation of the LHS}

The LHS covers about 28,000 square degrees, mostly of the Northern sky, to a
apparent magnitude limit of $R = 18.6$ and with a proper motion window of $0.5
< \mu < 2.5$ arcsec/year. We simulate the LHS by covering the fraction of the
sky ($\delta > -33^\circ$ and $|b| > 10^\circ$, i.e. 65\% of the sky) which is
estimated to be complete to better than 90\% by Dawson (1986).

\begin{figure}
\begin{center}
\epsfig{file=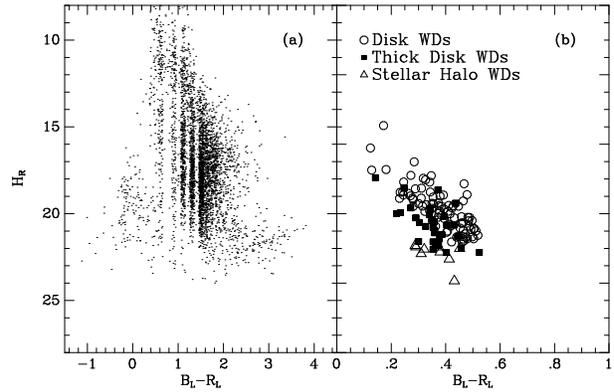,width=80mm}
\caption[]{Reduced proper motion versus colour for the LHS
sample. Panel (a) shows the data from the LHS, while panel (b) shows a
typical simulation (of the white dwarfs only). The white dwarf part of
the LHS data is broadly reproduced by the simulation.}
\label{lhscomp}
\end{center}
\end{figure}

Figure \ref{lhscomp} shows the reduced proper motion $H_R$ (RPM)
versus colour for the simulated LHS sample, where the Luyten $R$ band
magnitude ($R_L$) is related to the $V$ magnitude via (see Flynn et al
2001, appendix B)

\begin{displaymath}
R_L = V - 0.17 - 0.228 \times (B-V).
\end{displaymath}

Note that we have not shown the M dwarfs in the simulation, because we
were unable to obtain satisfactory transformations from $V-I$ to
Luyten's photographic $B_L-R_L$ colour.

Although the LHS and the OHDHS surveys probe similar volumes of space,
it is much easier to compare our simulations with OHDHS. This is
because the OHDHS sources have been spectroscopically classified into
white dwarfs and red dwarfs (such work is underway with the LHS and
should be available in the near future) and furthermore the colour
transformations are better understood than in the LHS.  Hence, we
regard the comparison of our model with OHDHS in the previous section
as superior to any comparison with the LHS. Nevertheless, the broad
features of the LHS observations are well reproduced by the
simulation, and we regard this as a qualitative consistency check on
the model.

\subsection{Dark Halo White Dwarfs?}

\begin{figure}
\begin{center}
\epsfig{file=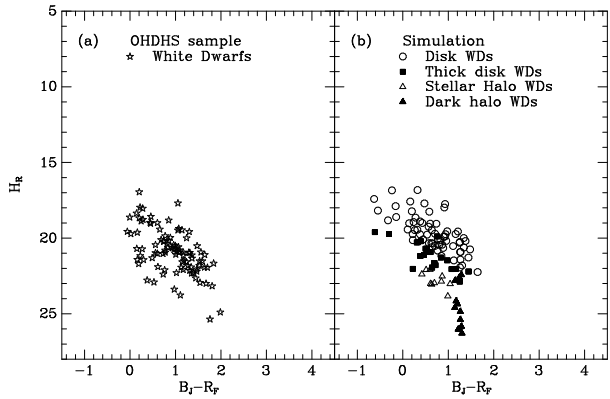,width=80mm}
\caption[]{Reduced proper motion versus colour plots for the OHDHS and
simulated samples. Panel (a) shows the data from OHDHS, with white dwarfs
marked by stars. Panel (b) shows a typical simulation of the OHDHS survey using
our model, now including white dwarfs from the dark halo (the figure is
otherwise the same as Figure \ref{oppcomp}).  Two percent of the dark halo mass
has been placed in white dwarfs of 0.6 $M_\odot$ at an absolute magnitude of
$M_V=15.9$, and these are shown in the simulated survey by filled triangles.
Most of the white dwarfs in OHDHS are plausibly from the disk, thick disk and
stellar halo, but two high reduced proper motion white dwarfs in OHDHS do lie
in the region where dark halo white dwarfs with an absolute magnitude of
$M_V=15.9$ are expected to lie, although (as shown in Figure \ref{frac}) the
simulations produce approximately a factor of ten more such stars than actually
seen in the combined LHS and OHDHS surveys.}
\label{Oppdarksim}
\end{center}
\end{figure}

We now introduce dark halo white dwarfs into the simulations. We begin
by adopting a luminosity function in which all the dark halo white
dwarfs are as faint as the faintest white dwarf found in the OHDHS
sample, i.e. at absolute magnitude $M_V = 15.9$. This choice {\it
minimises} the number of dark halo white dwarfs generated in the
simulations, while not being inconsistent with the luminosities of the
white dwarfs actually found in the OHDHS survey. Still, it appears to
produce too may simulated dark halo sources. Following OHDHS, we adopt
a white dwarf mass of 0.6 $M_\odot$ and local density of 2\% of the
dark halo, i.e. $0.02 \times 0.008 = 1.6 \times 10^{-4}$ \Msun
pc$^{-3}$.  In Figure \ref{Oppdarksim} we show where these dark halo
white dwarfs lie in the reduced proper motion versus colour plane.
Despite being conservative and minimising the number of ``dark'' white
dwarfs, there are plenty of these white dwarfs at a reduced proper
motion of $H_R \approx 25$, where just a few stars are found in the
observed samples.

\begin{figure}
\begin{center}
\epsfig{file=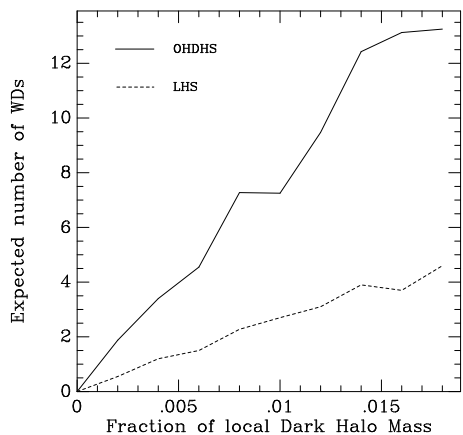,width=60mm}
\caption[]{Expected number of high reduced proper motion ($H_R > 24$) white
dwarfs in the two surveys analysed, as a function of the fraction of the local
dark matter density in such objects (for a white dwarf mass of $M_{\mathrm WD}
= 0.6 $M$_\odot$.  The actual number of objects found in both surveys with this
high a reduced proper motion is 2, which corresponds to a local density of
$\approx 0.2$\% of the dark halo. The dark halo density adopted is 0.008
M$_\odot$ pc$^{-3}$}
\label{frac}
\end{center}
\end{figure}

Judging from the simulations, a good discriminator between white
dwarfs of the disk, thick disk and stellar halo and those of the dark
halo is to divide them at $H_R = 24$. In the OHDHS sample, two white
dwarfs are found with reduced proper motion $H_R > 24$, while none
were found in the LHS. Both samples survey very similiar local
volumes. We plot in Figure \ref{frac} the number of high reduced
proper motion white dwarfs ($H_R > 24$) expected in the LHS and OHDHS
surveys as a function of their fraction of the dark halo. Assuming
that no dark halo white dwarf candidates were found in the LHS, and
two (the two absolutely faintest stars) were found in Oppenheimer et
al's survey, we conclude that approximately 0.4\% of the dark matter
density could be in white dwarfs (i.e. 2 WDs were found corresponding
to a dark matter fraction of $\approx 0.1$\% in figure \ref{frac}). A
dark matter fraction of 2\%, as suggested by OHDHS, would yield about
12 high reduced proper motion white dwarfs in the OHDHS sample and
about 5 in the LHS sample, for a total of 17, compared to 2 WDs
actually found.  Judging from both surveys, we conclude that OHDHS's
estimated fraction of 2\% of the dark halo density in white dwarfs is
an overestimate, and should be approximately an order of magnitude
smaller. If a significant fraction ($>0.1$\%) of the dark halo is in
white dwarfs, then they should be fainter than absolute magnitude $M_V
\approx 16$ in order to avoid being detected in large numbers in
either the LHS or OHDHS.

\section{Conclusions}

We have built a model of the kinematics and luminsities of low mass
stars in the Solar neighbourhood, which we use to simulate the results
of various proper motion surveys. We discuss in particular the search
for very faint white dwarfs in the Luyten Half Second proper motion
survey and recently by Oppenheimer et al (2001), which are the largest
surveys of the type. The surveys sample similar volumes of space. We
argue that the contribution of ``normal'' white dwarfs of the thick
disk and stellar halo is sufficient to explain the high velocity white
dwarfs found by OHDHS, and that they are not necessarily part of a
new, massive population from the dark halo. This work confirms results
from a similar study by Reyl\'e et al (2001).

\section*{Acknowledgments}

This research was supported by the Academy of Finland through the
ANTARES program for space research. Our thanks to Hugh Harris for much
help with the local halo white dwarfs in the LHS and Mike Bessell for
help with colour transformations. Burkhard Fuchs and Hartmut Jahrei\ss
are thanks for data on nearby subdwarf stars and many helpful
comments.

\end{document}